# Raman scattering of perovskite $DyScO_3$ and $GdScO_3$ single crystals

O. Chaix-Pluchery[1] and J. Kreisel

*Laboratoire des Matériaux et du Génie Physique,*
*CNRS UMR 5628 - Grenoble Institute of Technology - Minatec, 3, parvis Louis Néel, 38016 Grenoble, France*

**Abstract**

We report an investigation of $DyScO_3$ and $GdScO_3$ single crystals by Raman scattering in various scattering configurations and at various wavelengths. The Raman spectra are well-defined and the reported spectral signature together with the mode assignment sets the basis for the use of Raman scattering for the investigation of RE-scandates. The observed positions of Raman modes for $DyScO_3$ are for most bands in reasonable agreement with recent theoretical *ab initio* predictions of the vibrational spectrum for the same material. Further to the phonon signature, a luminescence signal is observed for both scandates. While the luminescence is weak for $DyScO_3$, it is very intense for $GdScO_3$ when using a 488 or 514 nm excitation line, which in turn inhibits full analysis of the phonon spectrum. We show that a meaningful phonon Raman analysis of $GdScO_3$ samples can be done by using a 633 nm excitation.

PACS 78.30.Hv

[1] to whom correspondence should be addressed (*E-mail: Odette.Chaix@inpg.fr*)

## 1. Introduction

Rare earth scandates (*RE*-scandates) with the generic formula $REScO_3$ receive currently an active research interest which is mainly focused on three issues: (*i*) *RE*-scandates are regarded as promising candidate materials for the replacement of $SiO_2$ in silicon MOSFETs, because they satisfy a number of stringent criteria [1-6]: a high dielectric constant *K*, chemical stability in contact with silicon, a large optical bandgap etc. Both crystalline and amorphous samples are investigated and the crystallization temperature is considered to be a key issue [6]. (*ii*) *RE*-scandate single crystals are considered to be among the best available substrates for the epitaxial growth of high-quality perovskite-type thin films as testified by their extremely narrow diffraction reflections [7]. Such high-quality films allow a strain engineering of ferroelectric [8-11] and multiferroic [10] properties by choosing different *RE*-scandate substrates: e.g. $SrTiO_3$, which is non-ferroelectric at any temperature, exhibits a strain-induced ferroelectricity when it is grown on *RE*-scandate substrates [8,10,12]. (*iii*) Scandates embedded in $SrTiO_3/DyScO_3$-type thin film heterostructures are currently investigated for applications in the terahertz range [13]. Advances in all these three issues depend on appropriate techniques for a detailed structural characterisation.

Raman spectroscopy, which probes zone-centre phonons, is known to be a versatile technique for the investigation of oxide materials in particular for the detection of even subtle structural distortions in perovskites [14-20]. For the particular case of thin films, Raman scattering has shown to be a powerful probe for the investigation of strain effects [21-23], texture [15], X-ray amorphous phases [17,24], heterostructure-related features [18,25,26], etc. Further to this, *experimentally* determined wavenumbers and symmetries of Raman phonons can be tested against *theoretically* predicted values. Such a comparison offers a meaningful test of *ab initio* calculations, complementary to the commonly used comparison of theory with diffraction-deduced structural data like lattice parameters.

*RE*-scandates crystallize is an orthorhombically distorted perovskite structure with space group *Pnma*. With respect to the ideal cubic *Pm3m* perovskite structure this orthorhombic structure is obtained by an anti-phase tilt of the adjacent $ScO_6$ octahedra ($a^-b^+a^-$ in Glazer's notation [27]). *RE*-scandates are thus ferroelastic, but we note that it is one of the particular advantages of the available *RE*-scandate substrates that they are grown without ferroelastic domains (contrary to $LaAlO_3$ or $BaTiO_3$ substrates). Raman spectra of distorted orthorhombic perovskite such as orthoferrites $REFeO_3$ [28,29], orthochromites

*RE*CrO$_3$ [30], orthomanganites *RE*MnO$_3$ [31], nickelates *RE*NiO$_3$ [16,32], etc., have been reported in the literature and contributed to their better understanding. On the other hand, to the best of our knowledge, orthoscandates *RE*ScO$_3$ have not yet been investigated by Raman scattering.

Here we present an experimental Raman scattering study of DyScO$_3$ and GdScO$_3$ with the aim (*i*) to provide reference data for future Raman investigations of bulk and thin film *RE*-scandates and (*ii*) to test the validity of the recently theoretically predicted *ab initio* vibrational spectrum of DyScO$_3$ [5].

## 2. Experimental and sample characterisation

We have investigated DyScO$_3$ and GdScO$_3$ single crystal platelets which are commercially available from *CrysTec* [33] with a size of 5*5*0.5 mm$^3$ in a (010) pseudo-cubic orientation. The starting materials for the crystal growth processes, Dy$_2$O$_3$, Gd$_2$O$_3$ and Sc$_2$O$_3$, were of 99.99% purity [34]. The straight edges of crystal platelets are parallel to the pseudo-cubic cell axes, i.e. to the [10-1] (= x') and [101] (= z') orthorhombic axes (figure 1), as confirmed by polarization properties of the Raman spectra. Thus, x and z orthorhombic axes are rotated in the (x'z') plane by 45° with respect to pseudo-cubic x' and z' axes, respectively (x = [100], z = [001]).

Raman spectra were recorded with a Jobin-Yvon/Horiba LabRam spectrometer equipped with a N$_2$-cooled CCD detector. Three wavelengths have been used: 488.0 and 514.5 nm lines of an Ar$^+$ ion laser and the 632.8 nm line of an He-Ne laser. Experiments were conducted in micro-Raman at room temperature; the light was focused to a 1$\mu$m$^2$ spot using a 50lf objective. All measurements performed under the microscope were recorded in a back-scattering geometry; the instrumental resolution was 2.8 ± 0.2 cm$^{-1}$. The Raman spectra of transparent DyScO$_3$ and GdScO$_3$ single crystals are not sensitive to the laser power and common experiments have been carried out with a laser power below 5 mW on the sample. Typical Raman spectra have been obtained for an acquisition time varying between 30 and 400 sec, depending on the exciting laser line and the polarization configuration. This illustrates that Raman scattering can be used as a rapid diagnostic tool for *RE*-scandates.

## 3. Results and discussion

The 20 atoms in the unit cell of the orthorhombic *Pnma* structure give rise to 24 Raman-active modes:

$$\Gamma_{Pnma} = 7A_g + 5B_{1g} + 7B_{2g} + 5B_{3g}$$

In the following we will first present the influence of the *RE*-atom (*RE* = Dy or Gd) on the *RE*-scandate Raman spectra; then, an assignment of the different symmetry modes will be derived from polarized Raman spectra of both compounds. Finally, the laser-wavelength-dependence of *RE*-scandate Raman spectra will be discussed.

*3.1. DyScO₃ and GdScO₃ Raman spectra*

We report in figure 2 Raman spectra of Dy- and Gd-scandates which have been obtained by using an excitation wavelength of 632.8 nm. At first sight, both spectra appear very similar below 550 cm$^{-1}$ concerning their number of lines and relative intensities. The line positions in both spectra are very close to each other especially at low wavenumbers and a closer comparison between phonon frequencies of both compounds indicate that they are usually lower for GdScO$_3$ (table 1). Generally speaking, small shifts in wavenumber in orthorhombic perovskites may originate either from the difference in rare earth atomic masses ($M_{Gd}$ = 157.25, $M_{Dy}$ = 162.5) or in their ionic radii ($r_{Gd}^{3+}$ = 1.053 Å, $r_{Dy}^{3+}$ = 1.027 Å, values are given for an eight-fold environment). It appears unlike that the very small difference in atomic mass between Gd$^{3+}$ and Dy$^{3+}$ can cause the observed shift and, furthermore, an inverse effect would be expected. On the other hand, it has been shown [35] that the ionic radius-dependence leads for RE-manganites to a linear frequency increase with decreasing $r_{RE}^{3+}$. Remind that decreasing $r_{RE}^{3+}$ leads to an increase of distortions through octahedra tilting and a shortening of bond lengths resulting in an increase of Raman frequencies. Based on this, we consider the ionic radius as the dominant factor in the variation of phonon frequencies in Gd- and DyScO$_3$. We finally note that the Raman spectra differ above 550 cm$^{-1}$, with the existence of broad lines on top of a very broad fluorescence band in the case of GdScO$_3$ (see discussion below).

*3.2. Symmetry assignment of Raman modes in DyScO₃ and GdScO₃*

Polarized Raman spectra of $DyScO_3$ and $GdScO_3$ are reported in figures 3a and 3b, respectively. As illustrated in the figures, our measurements have been performed in the following polarization configurations (Porto's notation) to separate the different symmetry modes:

$$y(xx)\bar{y}, \ y(zz)\bar{y}, \ y(x'x')\bar{y}, \ y(z'z')\bar{y} \ \rightarrow A_g,$$

$$y(xz)\bar{y}, \ y(x'x')\bar{y}, \ y(z'z')\bar{y} \ \rightarrow B_{2g},$$

$$z(xy)\bar{z}, \ x(yz)\bar{x} \ \rightarrow B_{1g} \text{ and } B_{3g}.$$

It is to be noted that we cannot distinguish x and z axes in our crystal platelets and therefore, neither are the polarization configurations such as *xx* and *zz*, *x'x'* and *z'z'*, *xy* and *zy*, respectively. Spectra given in the parallel $y(xx)\bar{y}$, $y(zz)\bar{y}$ and crossed $y(xz)\bar{y}$ polarization configurations in each figure allow assigning modes of $A_g$ and $B_{2g}$ symmetries, respectively. The 45° rotation of the samples from the $y(xx)\bar{y}$ or $y(zz)\bar{y}$ to the $y(x'x')\bar{y}$ *or* $y(z'z')\bar{y}$ configuration allows modes of both symmetries to be observed. Following these considerations, our spectral analysis leads to the following mode attribution: $A_g$ modes occur at 111, 131, 254, 327, 434, 459 and 509 cm$^{-1}$ in $DyScO_3$, at 113, 131, 248, 321, 418, 452 and 501 cm$^{-1}$ in $GdScO_3$; modes at 112, 156, 176, 309, 355, 475, 542 cm$^{-1}$ and the very weak line at 662 cm$^{-1}$ in $DyScO_3$ are of $B_{2g}$ symmetry; similarly $B_{2g}$ modes are observed at 115, 155, 174, 298, 351, 463, 532 cm$^{-1}$ in $GdScO_3$. The crossed $z(xy)\bar{z}$ or $x(yz)\bar{x}$ polarization configuration allows measuring the modes assigned as $B_{1g}$ or $B_{3g}$ but does not allow their experimental distinction in terms of symmetry.

Table 1 presents also a comparison of recent *ab initio* calculations of $DyScO_3$ vibrational modes at the zone-center [5] with our experimental Raman data. A reasonable overall agreement is observed between experimental and theoretical value for each symmetry mode except the additional mode observed at 176 cm$^{-1}$ (at 174 cm$^{-1}$ in $GdScO_3$). Although this mode obeys the selection rules, its origin and assignment remain to be elucidated. We note that experimental frequencies are always higher than calculated values. The correspondence between theoretical and experimental data then allows proposing the following assignment of the experimentally observed modes: the modes at 226, 381, 492 and 582 cm$^{-1}$ in $DyScO_3$ (223, 373, 490 and 573 cm$^{-1}$ in $GdScO_3$) are of $B_{1g}$ symmetry and those at 301, 457, 477 and 666 cm$^{-1}$ in $DyScO_3$ (300, 450, 481 and 669 cm$^{-1}$ in $GdScO_3$) are of $B_{3g}$ symmetry. The assignment of the two lines at 542 and 573 cm$^{-1}$ in $GdScO_3$ is not straightforward, but the comparison with $DyScO_3$ suggests that the 573 cm$^{-1}$ line is a $B_{1g}$ mode, while the 542 cm$^{-1}$ one is a fluorescence line. Unfortunately, the symmetry of the first mode (104 and 110 cm$^{-1}$ in Dy- and GdScO$_3$, respectively) cannot be discriminated between $B_{1g}$ and $B_{3g}$ because only one line is experimentally observed in comparison to two expected calculated lines in DyScO$_3$, $B_{3g}$ at 101.7 and $B_{1g}$ at 107.0 cm$^{-1}$ (see table 1). The upper lines at 626 and 690 cm$^{-1}$ in the GdScO$_3$ spectrum are probably fluorescence lines because they are not sensitive to polarization conditions; we will see below that experiments at different wavelengths validate this hypothesis.

*3.3. Influence of the exciting wavelength on DyScO$_3$ and GdScO$_3$ Raman spectra*

Raman spectra of Dy- and Gd-scandates collected by using three different exciting laser lines (488, 514.5 and 632.8 nm) are reported in figures 4a and 4b, respectively. We first note that the Raman signature of DyScO$_3$ is identical for all three investigated wavelengths with a very weak fluorescence signal at high wavenumbers (inset in figure 4a). In sharp contrast to this, the GdScO$_3$ spectra change significantly when changing the wavelength, which is a direct evidence of the presence of fluorescence lines. In the case of a 632.8 nm excitation, the GdScO$_3$ Raman phonon spectrum is only slightly disturbed by the presence of fluorescence lines above 550 cm$^{-1}$. On the other hand, when a 514.5 or 488 nm excitation is used, the GdScO$_3$ spectra are significantly disturbed from 200 or 350 cm$^{-1}$ towards higher wavenumbers by strong fluorescence lines, which make the phonon analysis considerably more complex, if not almost impossible, for these two wavelengths. The comparison of the three spectra below 700 cm$^{-1}$ in figure 4b shows that only eight common Raman lines are observed for the three wavelengths and that the fluorescence is less dominant when exciting with the 632.8 nm He-Ne laser line.

Therefore, it is not straightforward to determine for GdScO$_3$ which are the Raman phonon lines and which are the fluorescence lines, especially because both contributions are overlapping. This problem was only overcome by comparing the spectra of GdScO$_3$ with the undisturbed phonon spectrum of DyScO$_3$. It is only this comparison, which demonstrates that the 632.8 nm excitation is among the three wavelengths the only viable wavelength for the investigation of Raman-active phonons in GdScO$_3$.

As seen in figure 4b, the fluorescence signature in the GdScO$_3$ spectra appears either as a large intense band centered at very high wavenumbers when exciting with the red laser line, or as narrow intense lines above 1000 cm$^{-1}$ in the case of the green line, or as a mixture of broad (600-1000 cm$^{-1}$) and narrow (>1200 cm$^{-1}$) lines with the blue one. GdScO$_3$ spectra presented in a wavelength scale in figure 5 indicates still more clearly that fluorescence lines are superimposed with the Raman spectrum in the case of the green and blue lines while a more intense fluorescence band is centered above 725 nm when exciting with the red laser line, i.e. outside the Raman spectrum area. At present the origin of the different fluorescence lines remains to be elucidated, but it is likely related to spurious impurities contained in the starting binary oxides. In a general matter, and to the best of our knowledge, *RE*-scandates have not yet been investigated by fluorescence studies. The detailed analysis of the fluorescence signature is beyond the scope of this paper and calls for further specific investigations.

**4. Concluding remarks**

We have presented a Raman scattering investigation of DyScO$_3$ and GdScO$_3$ single crystals using three different excitations: 488, 514.5 and 632.8 nm. Among the 24 expected Raman-active phonon modes, 23 modes have been identified. A symmetry assignment has been proposed on the basis of the analysis of various polarisation configurations. The presented experimental reference data set the basis for the use of Raman scattering in DyScO$_3$ and GdScO$_3$, as for instance the investigation of strain (via phonon shifts) or texture (via polarised measurements) in thin films, but offers also the potential to follow the amorphous-crystalline state, etc.

The experimentally determined Raman mode wavenumbers for DyScO$_3$ are for most bands in reasonable agreement with values which have been recently reported by using theoretical *ab initio* calculations [5]. Reminding that phonons are a sensitive and versatile tool for testing *ab initio* models, our experiments suggest that the theoretically proposed cation charge anomalies and *high k* dielectric behaviour in DyScO$_3$ [5] are realistic.

The Raman spectrum of DyScO$_3$ samples can be investigated with all three wavelengths. In the other hand, we observe a significant fluorescence signal for GdScO$_3$ for all investigated wavelengths. In the case of a 488 or 514.5 nm excitation, the fluorescence lines are superimposed with the Raman signature inhibiting a meaningful analysis of the phonon spectrum. In contrast, the fluorescence of GdScO$_3$ using a 632.8 nm excitation is outside the phonon range and thus, it can and should be used for a Raman analysis of GdScO$_3$ samples. However, it is worth noting that the fluorescence should not only be regarded as a source of problems for Raman investigations, but might also turn out to be a versatile local probe for phase transitions in *RE*-scandates, complementary to Raman and XRD as has been shown for other materials [35]. We hope that the presented data motivate other research groups to use Raman scattering as a structural probe in scandates (just as is often done in other oxides) and that the origin and symmetry assignment of the fluorescence signal of GdScO$_3$ attracts further attention.

*Acknowledgements*

This work was supported by the European Network of Excellence FAME (Functionalized Advanced Materials and Engineering) and the European Strep MaCoMuFi.

**Figure captions**

Figure 1: Schematic representation of the orientation of the $DyScO_3$ and $GdScO_3$ crystal platelets in the orthorhombic unit cell.

Figure 2: Comparison between depolarized Raman spectra of $DyScO_3$ and $GdScO_3$ collected with using the 632.8 nm He-Ne laser line. The propagation directions of the incoming and collected photons are along the $y$ axis of the crystal platelets.

Figure 3: Polarized Raman spectra of $DyScO_3$ (a) and $GdScO_3$ (b) collected in various polarization configurations with using the 632.8 nm He-Ne laser line. The positions of phonon modes allowed in each configuration are given in bold [$A_g$ in $y(xx)\bar{y}$ or $y(zz)\bar{y}$, $B_{2g}$ in $y(xz)\bar{y}$, $B_{1g}$ and $B_{3g}$ in $z(xy)\bar{z}$ or $x(yz)\bar{x}$].

Figure 4: Polarized Raman spectra of $DyScO_3$ (a) and $GdScO_3$ (b) collected in the $y(x'x')\bar{y}$ or $y(z'z')\bar{y}$ polarization configuration as a function of the exciting laser line (632.8, 514.5 and 488 nm). The intensity of the high wavenumber part of the $DyScO_3$ spectra shown in the inset is multiplied by a factor 3. Arrows denote the eight Raman lines common in the three $GdScO_3$ spectra.

Figure 5: $GdScO_3$ spectra of figure 4b presented in a wavelength scale (nm). The Raman phonon area of the spectra defined inside dashed rectangles provide evidence that fluorescence lines are superimposed to the Raman signature in the case of 488 and 514.5 nm exciting lines while a large fluorescence signal occurs outside the Raman spectrum area when exciting with the 632.8 nm laser line.

Table 1: Phonon wavenumbers experimentally obtained from polarized Raman spectra of DyScO$_3$ and GdScO$_3$ single crystals. Calculated values reported in reference [5] for DyScO$_3$ are also given for comparison.

| Symmetry mode | DyScO$_3$ $\omega_{calc}$ (cm$^{-1}$) | DyScO$_3$ $\omega_{exp}$ (cm$^{-1}$) | GdScO$_3$ $\omega_{exp}$ (cm$^{-1}$) | DyScO$_3$ $\Delta\omega$ (exp-calc) (%) | $\Delta\omega_{exp}$ (DyScO$_3$ - GdScO$_3$) (%) |
|---|---|---|---|---|---|
| A$_g$ | 108.4 | 111 | 113 | 2.8 | -1.8 |
| A$_g$ | 119.7 | 131 | 131 | 9.2 | 0 |
| A$_g$ | 243.7 | 254 | 248 | 4.1 | 2.4 |
| A$_g$ | 319.6 | 327 | 321 | 2.2 | 1.9 |
| A$_g$ | 417.9 | 434 | 418 | 3.8 | 3.8 |
| A$_g$ | 441.4 | 459 | 452 | 4.1 | 1.5 |
| A$_g$ | 485.3 | 509 | 501 | 4.9 | 1.6 |
| B$_{2g}$ | 108.6 | 112 | 115 | 3.1 | -2.6 |
| B$_{2g}$ | 153.5 | 156 | 155 | 1.6 | 0.6 |
| ? | ? | 176 | 174 | - | 1.1 |
| B$_{2g}$ | 300.6 | 309 | 298 | 2.8 | 3.7 |
| B$_{2g}$ | 340.9 | 355 | 351 | 4.1 | 1.1 |
| B$_{2g}$ | 461.0 | 475 | 463 | 3.0 | 2.6 |
| B$_{2g}$ | 518.9 | 542 | 532 | 4.5 | 1.9 |
| B$_{2g}$ | 630.2 | 662 | - | 5.0 | - |
| B$_{3g}$ | 101.7 | 104 ? | 110 ? | 2.3 | - |
| B$_{3g}$ | 286.7 | 301 | 300 | 5.0 | 0.3 |
| B$_{3g}$ | 432.9 | 457 | 450 | 5.6 | 1.6 |
| B$_{3g}$ | 442.6 | 477 | 481 | 7.7 | -0.8 |
| B$_{3g}$ | 634.6 | 666 | 669 | 4.9 | -0.4 |
| B$_{1g}$ | 107.0 | 104 ? | 110 ? | - | - |
| B$_{1g}$ | 220.3 | 226 | 223 | 2.6 | 1.3 |
| B$_{1g}$ | 363.4 | 381 | 373 | 4.8 | 2.1 |
| B$_{1g}$ | 461.1 | 492 | 490 | 6.7 | 0.4 |
| B$_{1g}$ | 573.9 | 582 | 573 | 1.4 | 1.6 |

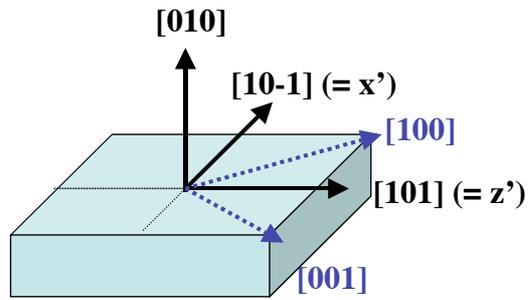

Figure 1

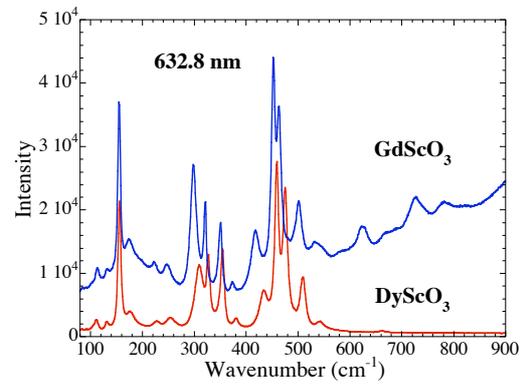

Figure 2

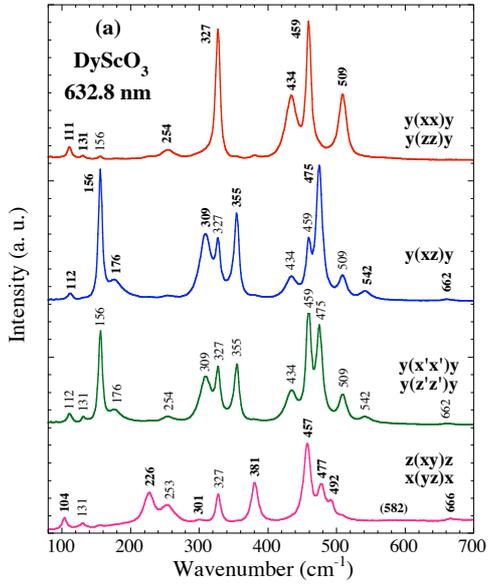 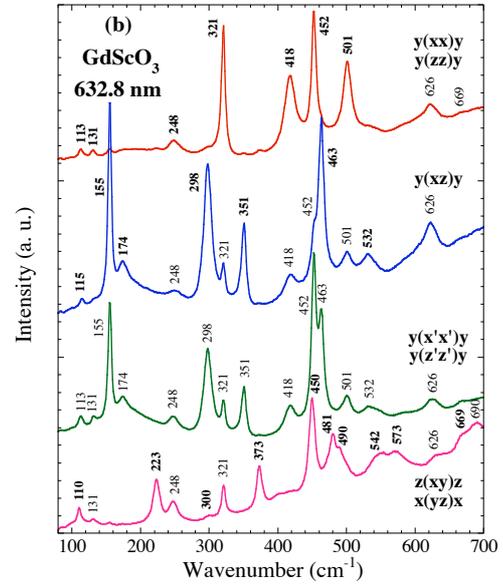

Figure 3a        Figure 3b

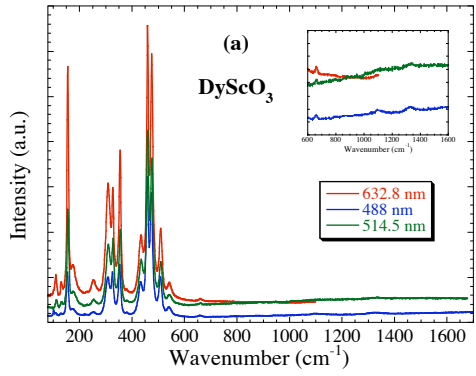
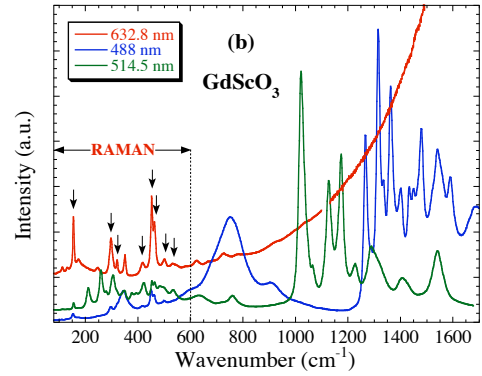

Figure 4a                    Figure 4b

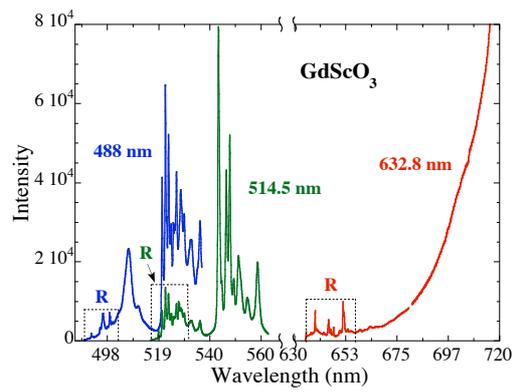

Figure 5